\documentstyle[eqsecnum,preprint,aps,epsf]{revtex}

\begin{document}

\hoffset -1cm

\draft

\preprint{TPR-95-16}

\title{Spectral functions for composite fields\\
       and viscosity in hot scalar field theory}

\author{Enke Wang\cite{address} and Ulrich Heinz}

\address{Institut f\"ur Theoretische Physik, Universit\"at Regensburg, \\
         D-93040 Regensburg, Germany}

\author{Xiaofei Zhang}

\address{Institute of Particle Physics, Hua-Zhong Normal University,\\
         Wuhan 430070, China}

\date{\today}

\maketitle

\begin{abstract}
 We derive a spectral representation for the two-point Green function
 for arbitrary composite field operators in Thermo Field Dynamics
 (TFD). A simple way for calculating the spectral density within TFD
 is pointed out and compared with known results from the imaginary
 time formalism. The method is applied to hot $\phi^4$ theory. We give
 a compact derivation of the one-loop contribution to the shear
 viscosity and show that it is dominated by low-momentum plasmons.
\end{abstract}


\narrowtext


\section{Introduction}\label{sec1}

In the context of heavy-ion collisions at ultrarelativistic energies,
transport properties of strongly interacting hot matter, in particular
of the quark-gluon plasma (QGP), have become an issue of intense
theoretical interest. The calculation of QGP transport coefficients
from linear response functions through the Kubo-formalism
\cite{Hosoya,Ilyin} has so far been hampered by infrared problems
associated with the non-linear selfinteractions among massless fields
in hot QCD. A systematic and economic field theoretical approach,
including resummation techniques for infrared divergences
\cite{Pisarski}, to the calculation of linear response functions is
clearly asked for.

As an example for the type of objects one is interested in let us take
the Kubo-formula for the shear viscosity \cite{Hosoya,Zubarev,Jeon}
  \begin{equation}
  \label{eta}
  \eta = -{1\over 5}\int d^3 x'\int_{-\infty}^0 dt
       \int_{-\infty}^{t} dt'\,
       \langle\pi ^{\mu\nu}(0),
       \pi _{\mu\nu}({\bf x}',t')\rangle^{\rm ret}
     = {\pi\over 5}\lim_{\omega,{\bf p} \to 0}
       {{\partial\rho_{\pi\pi}(\omega, {\bf p})}\over {\partial\omega}}
       \, .
  \end{equation}
Here $\pi _{\mu\nu}$ is the traceless viscous-pressure tensor which is
imbedded in the energy-momentum tensor and in the comoving frame has
only spatial components, $\pi_{ij}$. For a scalar field $\pi_{ij} =
(\delta_{ik}\delta_{jl}-{1\over 3}\delta_{ij}\delta_{kl}) \partial_k
\phi\,\partial_l\phi$. $\rho_{\pi\pi}$ is the spectral momentum space
density for the retarded thermal Green function of the composite field
$\pi$, as defined below. The efficient computation of this quantity
will be the subject of this paper.

The analytical structure of the retarded Green function $\langle
\pi^{\mu\nu} (0), \pi _{\mu\nu}({\bf x}',t')\rangle^{\rm ret}$ is similar to
that of $\langle\phi^2(0), \phi^2({\bf x}',t')\rangle^{\rm ret}$ \cite{Jeon}.
The spectral density for the latter 2-point function was recently
calculated at 1-loop order in the imaginary time formalism (ITF)
by Jeon \cite{Jeon}. Here we repeat the calculation in the real-time
formalism using the language of Thermo Field Dynamics (TFD)
\cite{Umezawa}. Using the analyticity properties of the 2-point
functions in ITF and TFD, we give a very compact derivation of the
1-loop contribution to the spectral density of the composite field
2-point function, $\rho_{\pi\pi}$, in terms of the spectral density of
the elementary field 2-point function, $\rho_{\phi\phi}$. The latter
was recently calculated at 2-loop order including infrared resummation
\cite{Jeon,Parwani,Wang}. Our approach follows closely the methods
developed in \cite{Kobes,Eijck,Evans} for correlation functions
between elementary fields. It goes beyond the results reported in
\cite{Jeon} in that we include the full momentum dependence of the
2-loop scalar self energy.

\section{Spectral representation of composite field propagators in
TFD}\label{sec2}

For two arbitrary operators $\hat A$, $\hat B$ the retarded thermal
Green function is defined by
  \begin{equation}
  \label{drab1}
    D^{\rm ret}_{_{\!AB}}({\bf r}_{1}-{\bf r}_{2},t_{1}-t_{2})
        =-i\theta (t_1-t_2)\, {\rm Tr} \Bigl( e^{\beta(\Omega-{\hat H})}
        [{\hat A}({\bf r}_{1},t_{1}),
                {\hat B}({\bf r}_{2},t_{2})]_{\pm} \Bigr) \, ,
  \end{equation}
where $\hat H$ is the hamiltonian, $\Omega$ is the thermodynamic
potential, and the plus (minus) sign applies for fermionic (bosonic)
operators $\hat A$, $\hat B$. It can be expressed as an integral over
the spectral density $\rho_{_{\!AB}}(\omega,{\bf p})$ in momentum space:
  \begin{equation}
  \label{drab2}
  D^{\rm ret}_{_{\!AB}}(p_0,{\bf p}) = \lim_{\eta\to 0^+}
     \int_{-\infty}^{+\infty} d\omega\,
     {\rho_{_{\!AB}}(\omega, {\bf p}) \over p_{0}-\omega+i\eta} \, .
  \end{equation}
In the Heisenberg picture the spectral density has the Lehmann
representation \cite{Lehmann}
  \begin{eqnarray}
  \label{rho}
     \rho_{_{\!AB}}(\omega,{\bf p})
     &=& (2\pi)^3 \sum_{n,m} e^{\beta(\Omega-E_n)}
  \nonumber\\
    &&\langle n\vert A(0)\vert m\rangle
     \langle m\vert B(0)\vert n\rangle \, (1-\sigma e^{-\beta\omega})
     \, \delta(\omega-E_{mn}) \, \delta^{(3)}({\bf p}-{\bf p}_{mn})\, ,
  \end{eqnarray}
where $\sigma=+1$ ($-1$) for bosonic (fermionic) operators $\hat A$,
$\hat B$, respectively. The sum is over a complete set of simultaneous
eigenvectors of $\hat H$ and $\hat {\bf p}$, with eigenvalues $E_n$
and ${\bf p}_n$, and $E_{mn}=E_{m}-E_{n}$, ${\bf p}_{mn}={\bf p}_{m}-
{\bf p}_{n}$.

The imaginary time Green function is obtained from Eq.~(\ref{drab1})
by setting $it=\tau$. In momentum space it has a similar spectral
representation,
 \begin{equation}
 \label{gab}
  G_{_{\!AB}}(\omega_n,{\bf p})=\int_{-\infty}^{+\infty} d\omega \,
     {\rho_{_{\!AB}}(\omega, {\bf p})\over i\omega_n-\omega} \, ,
  \end{equation}
in terms of the same spectral function $\rho_{_{\!AB}}$. The periodicity of
the ITF Green functions in imaginary time results in the discrete
Matsubara frequencies $\omega_n=2n\pi T$ and $(2n+1)\pi T$ for bosonic
and fermionic operators $\hat A$ and $\hat B$, respectively. The
retarded Green function (\ref{drab2}) and the ITF Green function
(\ref{gab}) are related by analytical continuation $i\omega_n\to
p_0+i\eta$:
  \begin{equation}
  \label{con}
  D^{\rm ret}_{_{\!AB}}(p_{0},{\bf p})=G_{_{\!AB}}(\omega_n,{\bf p})
    \Big\vert_{i\omega_n\to p_0+i\eta}\, .
  \end{equation}
The propagators in TFD have a more complicated structure:
any physical field operator $\hat A$ is associated with
a thermal ghost field operator $\tilde A$, with which it is combined
into a doublet:
  \begin{equation}
  \label{A12}
  {A^{1}\choose A^{2}}={{\hat A}\choose {\tilde A}^{\dagger}}\, .
  \end{equation}
The TFD Green function is thus a $2\times 2$ matrix
  \begin{equation}
  \label{d1}
  \Delta_{_{\!AB}}^{ab}({\bf r}_{1}-{\bf r}_{2},t_{1}-t_{2})
        = -i\, \langle 0,\beta\vert T_{t}[{\hat A}^{a}({\bf r}_{1},t_{1}),
          {\hat B}^{b}({\bf r}_{2},
          t_{2})]_{\pm}\vert 0,\beta\rangle \, , \qquad
  a,b=1,2\, .
  \end{equation}
In momentum space it has the following spectral representation
  \begin{equation}
  \label{d2}
    \Delta_{_{\!AB}}^{ab}(p_{0},{\bf p}) =
    \int_{-\infty}^{+\infty} d\omega\,
    \left[ {\gamma_{_{\!AB}}^{ab}(\omega,{\bf p})
            \over p_0-\omega+i\eta}
          - \sigma {\gamma_{_{\!BA}}^{ba} (\omega,-{\bf p})
            \over p_0+\omega-i\eta}
    \right]\, ,
  \end{equation}
where $T_t$ is time ordering and the spectral function has the
generalized Lehmann representation
  \begin{eqnarray}
  \label{gamma}
    \!\!\!\!\!\!\!\! \gamma_{_{\!AB}}^{ab}(\omega,{\bf p})
    \!\! &=&\!\! (2\pi)^3 \sum_{n,m}
        \langle 0,\beta\vert A^{a}(0) \vert n,{\tilde m}\rangle
        \langle{\tilde m},n\vert B^{b}(0)\vert 0,\beta\rangle \,
        \delta(\omega-E_{nm}) \, \delta^{(3)}({\bf p}-{\bf p}_{nm})\, ,
  \\
  \label{gamma+1}
    \!\!\!\!\!\!\!\! \gamma_{_{\!BA}}^{ba}(\omega,-{\bf p})
    \!\!&=&\!\! (2\pi)^3 \sum_{n,m}
        \langle 0,\beta\vert B^{b}(0) \vert n,{\tilde m}\rangle
        \langle{\tilde m},n\vert A^{a}(0)\vert 0,\beta\rangle \,
        \delta(\omega-E_{nm}) \, \delta^{(3)}({\bf p}+{\bf p}_{nm})
    \, .
  \end{eqnarray}
Here the thermal vacuum state $\vert 0,\beta\rangle$ is defined
as \cite{Umezawa}
  \begin{equation}
  \label{vacumm}
  \vert 0,\beta\rangle=\sum_k e^{\beta(\Omega- E_k)/2}
        \vert k,{\tilde k}\rangle\, ,
  \end{equation}
where $\vert k,{\tilde k}\rangle=\vert k\rangle\otimes
\vert {\tilde k}\rangle$, and $\vert {\tilde k}\rangle$ is eigenstate
of $\tilde H$ with the same eigenvalue $E_k$.

Using Eq.~(\ref{vacumm}) and the `tilde substitution law' \cite{Umezawa}
  \begin{eqnarray}
  \label{subslaw}
    \langle 0, \beta\vert  {\tilde A}^{\dagger}(0)\vert n,{\tilde m}\rangle
    &=&(-1)^{F_A(F_A-1)/2}\exp[-\beta E_{nm}/2]
    \langle 0, \beta\vert  {\hat A}(0)\vert n,{\tilde m}\rangle\, ,
  \\
  \label{subslaw1}
    \langle {\tilde m},n \vert {\tilde  A}^{\dagger}(0)\vert 0,\beta\rangle
    &=&(-1)^{F_A(F_A+1)/2}\exp[-\beta E_{nm}/2]
    \langle {\tilde m},n \vert {\hat  A}(0)\vert 0,\beta\rangle\, ,
  \end{eqnarray}
where $F_A$ is the fermion number associated with the operator $\hat
A$, and noticing that the physical field $\hat A$ only acts on
the physical states $\vert n\rangle$ while the ghost field $\tilde A$
acts only on the tilde states $\vert {\tilde n}\rangle$,
  \begin{eqnarray}
  \label{pg}
  \langle {\tilde k},l \vert  {\hat A}\vert n,{\tilde m}\rangle
     &=&\delta_{km} \, \langle l\vert {\hat A} \vert n\rangle\, ,
  \\
  \label{pg1}
  \langle {\tilde k},l \vert {\tilde  A}\vert n,{\tilde m}\rangle
     &=&\delta_{ln} \, \langle {\tilde k}\vert {\tilde  A}
     \vert {\tilde m}\rangle\, ,
  \end{eqnarray}
we can transform $\gamma_{_{\!AB}}^{ab}$ in Eqs.~(\ref{gamma}) and
(\ref{gamma+1}) into similar form as Eq.~(\ref{rho}). Some further
algebra involving Eqs.~(\ref{drab2}) and (\ref{d2}) then yields for
the TFD propagator matrix the result
  \begin{eqnarray}
  \label{d3}
    \lefteqn{ \Delta_{_{\!AB}}(p_0,{\bf p})=}
  \\
    &&\!\!\!\!\!\!\!\!\!\!\!\!\!\!
      \left( \begin{array}{ccc}
       D^{\rm ret}_{_{\!AB}}(p_{0},{\bf p})
       - {\displaystyle
          { 2\pi i \, \sigma \over \exp(\beta p_{0}) - \sigma } }
         \rho_{_{\!AB}}(p_{0},{\bf p}) \,,
    &
       -2\pi i\,\sigma
        {\displaystyle
         {s_{_{\!AB}} \exp(\beta p_{0}/2) \over \exp(\beta p_{0})-\sigma} }
         \rho_{_{\!AB}}(p_{0},{\bf p})
    \\
    \quad\quad\\
       -2\pi i\,\sigma
        {\displaystyle
         {s_{_{\!AB}} \exp(\beta p_{0}/2) \over \exp(\beta p_{0})-\sigma }}
        \rho_{_{\!AB}}(p_{0},{\bf p})\, ,
    &
       - \sigma \left(D^{\rm ret}_{_{\!AB}}(p_{0},{\bf p})\right)^*
       - {\displaystyle
          { 2\pi i \over \exp(\beta p_{0})-\sigma}}
         \rho_{_{\!AB}}(p_{0},{\bf p})
  \end{array} \right),
  \nonumber
  \end{eqnarray}
where $s_{_{\!AB}} = (-1)^{F_A(F_A+1)/2}$, using $F_A=-F_B$ since
otherwise $\Delta_{_{\!AB}}$ vanishes \cite{Umezawa}.

With (\ref{drab2}) this provides explicit expressions for each
component of the 2-point Green function in terms of the spectral
density $\rho_{_{\!AB}}$. The relationship of $\rho_{_{\!AB}}$ to the
(12)- and (21)-components of $\Delta_{_{\!AB}}$ is particularly
straightforward:
  \begin{equation}
  \label{rho1}
  \rho_{_{\!AB}}(p_0,{\bf p}) = s_{_{\!AB}}\,{ i\sigma \over 2\pi} \,
       { e^{\beta p_0}-\sigma \over e^{\beta p_0/2} } \,
       \Delta^{12}_{_{\!AB}}(p_0,{\bf p})\, .
  \end{equation}
Since the (12)-component of the 2-point Green function can be
calculated directly using TFD Feynman rules, this relation provides a
convenient way of obtaining the spectral density for the composite
fields $\hat A$, $\hat B$.

The above expressions apply in particular to the single-particle
operators ${\hat A}=\phi$ and ${\hat B}=\phi^{\dagger}$ ($F_{_A} = -
F_{_B} = 0$, $\sigma=1$). As shown in \cite{Jeon}, $\rho_{\phi
\phi^{\dagger}} (\omega,{\bf p})$ is real, depends only on $\vert {\bf
p}\vert$ due to spatial isotropy, and is for any CPT-invariant
equilibrium state an odd function of the frequency,
  \begin{equation}
  \label{odd}
  \rho_{\phi\phi^{\dagger}}(\omega,{\bf p})=
    \epsilon(\omega)\rho_{\phi\phi^{\dagger}}(|\omega|,|{\bf p}|)\, ,
  \end{equation}
where $\epsilon(\omega)={\rm sgn}(\omega)$. Using Eq.~(\ref{con}) and
the following relation \cite{Weldon} for the analytically continued
ITF propagator
  \begin{equation}
  \label{odd1}
   G_{\phi\phi^{\dagger}}(p_0+i\eta, {\bf p})=
   G_{\phi\phi^{\dagger}}(p_0+i\epsilon(p_0)\eta, {\bf p})+2\pi i\theta(-p_0)
   \rho_{\phi\phi^{\dagger}}(|p_0|,{\bf p})\, ,
  \end{equation}
we can write the full TFD propagator for the scalar field as
  \begin{eqnarray}
  \label{propf}
  \lefteqn{ \Delta\equiv\Delta_{\phi\phi^{\dagger}}(p_0,{\bf p})=}
  \\
    & &\!\!\!\!\!\!\!\!\!\!\!\!\!
    \left( \begin{array}{ccc}
     G_{\phi\phi^{\dagger}}(p_0+i\epsilon(p_0)\eta, {\bf p})
    -{\displaystyle
      {2\pi i\,\rho_{\phi\phi^{\dagger}} (|p_{0}|,{\bf p})
       \over \exp(\beta |p_0|) - 1 }} \, , &
      -2\pi i\,
      {\displaystyle
       { \exp(\beta |p_0|/2) \over
         \exp(\beta |p_0|) - 1 }}\,
      \rho_{\phi\phi^{\dagger}}(|p_{0}|,{\bf p}) \\
    \quad\quad\\
      -2\pi i\,
      {\displaystyle
       { \exp(\beta |p_0|/2) \over
         \exp(\beta |p_0|) - 1}}\,
      \rho_{\phi\phi^{\dagger}}(|p_{0}|,{\bf p})\, , &
     -G^*_{\phi\phi^{\dagger}}(p_0+i\epsilon(p_0)\eta, {\bf p})
     -{\displaystyle
       { 2\pi i \, \rho_{\phi\phi^{\dagger}}(|p_0|,{\bf p})
         \over \exp(\beta |p_0|) - 1} }
   \end{array} \right) \, ,
  \nonumber
  \end{eqnarray}
with the following representations in terms of the ITF self energy
$\Sigma$:
  \begin{eqnarray}
  \label{gphi}
   G_{\phi\phi^{\dagger}}(p_0+i\epsilon(p_0)\eta,{\bf p})
   &=& {1 \over p^2 - m^2 + {\rm Re\,}\Sigma(|p_0|,{\bf p})
              + i\, {\rm Im\,}\Sigma(|p_0|,{\bf p})}\, ,
  \\
  \rho_{\phi\phi^{\dagger}}(|p_0|,{\bf p})
  &=& -{1\over 2\pi i}\,
      \Bigl[ G_{\phi\phi^{\dagger}}(|p_0|+i\eta,{\bf p})
           - G_{\phi\phi^{\dagger}}(|p_0|-i\eta,{\bf p}) \Bigr]
  \nonumber\\
    &=& {1\over \pi}\,
        { {\rm Im\,}\Sigma(|p_0|,{\bf p})
          \over
          \bigl[ p^2-m^2+{\rm Re\,}\Sigma(|p_0|,{\bf p})\bigr]^2
         +\bigl[ {\rm Im\,}\Sigma(|p_0|,{\bf p})\bigr]^2 }\, .
  \label{rhof}
  \end{eqnarray}
For a free scalar field both real and imaginary parts of $\Sigma$
vanish, and the spectral density $\rho_{\phi\phi^{\dagger}}$
degenerates to a $\delta$-function:
  \begin{equation}
  \label{da}
    {1\over \pi} \lim_{\eta \to 0}{\eta\over {p^2-m^2+\eta^2}}
    =\delta(p^2-m^2)\, .
  \end{equation}
Eq.(\ref{propf}) in this case reduces to the free TFD propagator
  \begin{equation}
  \label{prop0}
  \Delta_0=
  \left( \begin{array}{ccc}
    {\displaystyle
     {1\over p^2-m^2+i\eta} - {2\pi i\,\delta(p^2-m^2)
      \over \exp(\beta |p_0|)-1}
     }, &
     {\displaystyle
      -{ 2\pi i\,\delta(p^2-m^2)\,\exp(\beta |p_0|/2)
         \over
         \exp(\beta |p_0|)-1 }
     } \\
     \quad\quad\\
     {\displaystyle
     -{2\pi i\,\delta(p^2-m^2)\,\exp(\beta |p_{0}|/2)
       \over
       \exp(\beta |p_{0}|)-1}
     }, &
     {\displaystyle
     -{1 \over p^2-m^2-i\eta}
     -{2\pi i\,\delta(p^2-m^2) \over \exp(\beta |p_0|)-1 }
     }
  \end{array} \right)\, .
  \end{equation}

The full TFD propagator matrix $\Delta$ satisfies the Dyson equation
\cite{Matsumoto}
  \begin{equation}
  \label{dyson}
  (\Delta^{-1})^{ab}=(p^2-m^2)\tau^{ab}
    +\Sigma^{ab}(p_0,{\bf p})\, ,
  \end{equation}
where $\tau={\rm diag}(1,-1)$. This defines a $2\times 2$ TFD self
energy matrix $\Sigma^{ab}$. Comparing this with the inverse of the
matrix (\ref{propf}), it is easy to obtain the following relations:
  \begin{eqnarray}
  \label{sele}
    {\rm Im\,}\Sigma
    &=& {e^{\beta p_0}-1\over e^{\beta p_0}+ 1}\, {\rm Im\,}\Sigma^{11}
      = -{e^{\beta p_0}-1\over 2e^{{\beta p_0}/ 2}}\,
         {\rm Im\,}\Sigma^{12} ,\ \
  \\
     {\rm Im\,}\Sigma^{21} &=& {\rm Im\,}\Sigma^{12},\ \
     {\rm Im\,}\Sigma^{22}  =  {\rm Im}\Sigma^{11},
  \\
  \label{sele1}
    {\rm Re\,}\Sigma
    &=& {\rm Re\,} \Sigma^{11} = -{\rm Re\,} \Sigma^{22}\, ,
   \quad
    {\rm Re\,} \Sigma^{12} = {\rm Re\,} \Sigma^{21}=0\, .
  \end{eqnarray}
They were previously obtained in \cite{Kobes,Fuji} by different
methods. The authors of Ref.~\cite{Fuji} pointed out that the
calculation of ${\rm Im\,}\Sigma$ from ${\rm Im\,}\Sigma^{12}$, using
TFD Feynman rules for the evaluation of the latter, is particularly
convenient and showed explicitly that the result is identical to the
evaluation in ITF.

It is straightforward to generalize the expressions (\ref{propf}) and
(\ref{prop0}) as well as the self energy relationships
(\ref{sele} - \ref{sele1}) to the case of fermionic single particle
operators.

\section{Shear viscosity in hot $\phi^4$ theory}\label{sec3}

We will now use these results to give a five line derivation of the
spectral function for the composite field $\phi^2$ within hot
$\lambda\phi^4$ theory. We concentrate on the same quantity as studied
in Ref.~\cite{Jeon}, namely
  \begin{equation}
  \label{eta1}
  \eta_{\phi^2\phi^2} \equiv \lim_{{\bf p},p_0\to 0}
       \left[ {\rho_{\phi^2\phi^2}(p_0, {\bf p}) \over p_0}\right]
       \, ,
  \end{equation}
whose relation to the shear viscosity (\ref{eta}) was mentioned above.
{}From Eq.~(\ref{rho1}) we obtain
  \begin{equation}
  \label{rho2}
  \rho_{\phi^2\phi^2}(p_{0},{\bf p})=
       {i\over {2\pi}}{{e^{\beta p_0}-1}\over e^{{\beta p_0}/2}}
       \Delta^{12}_{\phi^2\phi^2}(p_0,{\bf p})\, .
  \end{equation}
The lowest order contribution to $\Delta^{12}_{\phi^2\phi^2}(p_0,{\bf
p})$ is shown diagrammatically in Fig.~1. This is a skeleton
diagram, i.e. the full single-particle propagators (\ref{propf}) must
be used for the internal lines \cite{Hosoya,Jeon}. For the one-loop
diagram in Fig.~1 only its (12)-component is needed. We find
  \begin{eqnarray}
  \label{d12}
  i \, \Delta^{12}_{\phi^2\phi^2}(p_0,{\bf p}) &=&
    2\int{{d^4k}\over {(2\pi)^2}}
    \left[ { \exp({\beta k_0\over 2}) \,\rho_{\phi\phi}(k_{0},{\bf k})
             \over
             \exp({\beta k_{0}})-1 } \right]
    \left[ { \exp({ \beta (p_0-k_0) \over 2}) \,
             \rho_{\phi\phi}(p_0-k_0,{\bf p-k})
             \over
             \exp(\beta (p_0-k_0))-1 } \right]
 \nonumber\\
  &=& 2\,e^{\beta p_0\over 2} \int {d^4k \over (2\pi)^2}\,
      f(k_0) \, f(p_0-k_0) \, \rho_{\phi\phi}(k_{0},{\bf k}) \,
      \rho_{\phi\phi}(p_0-k_{0},{\bf p-k})\, ,
  \end{eqnarray}
with the Bose distribution
  \begin{equation}
  \label{dis}
  f(k_0)={1\over {\exp(\beta k_0)-1}}\, .
  \end{equation}
Combining this with Eqs.~(\ref{rho2}) and (\ref{eta1}) we obtain
  \begin{equation}
  \label{eta2}
  \eta^{{\rm 1-loop}}_{\phi^2\phi^2} =
    2 \beta \int {d^4k\over (2\pi)^3}\,
    f(k_0) \, [1+f(k_0)] \, [\rho_{\phi\phi}(k_{0},|{\bf k}|)]^2\, ,
  \end{equation}
where we used Eq.~(\ref{odd}) as well as the identity $f(-k_0) = -
[1+f(k_0)]$. The factor $\beta$ results from ${\displaystyle
\lim_{p_0\to 0}} (e^{\beta p_0}-1)/p_0$. Up to a factor of $2\pi$
resulting from our different normalization (\ref{rhof}) of the
spectral density, this result is identical with the (much lengthier)
calculation \cite{Jeon} from the cutting rules in ITF.

To further evaluate Eq.~(\ref{eta2}) we need the spectral density
$\rho_{\phi\phi}(k_{0},{\bf k})$ for the full single particle
propagator. Since the calculation of $\eta^{{\rm 1-loop}}_{\phi^2
\phi^2}$ requires taking the zero momentum limit of the loop diagram,
both internal lines in Fig.~1 can become soft. For massless
$\lambda \phi^4$ theory this means that, in order to avoid infrared
divergences, we should use resummed effective propagators
\cite{Pisarski,Parwani}. The so-called resummation of ``hard thermal
loops'' \cite{Pisarski} in this case generates a thermal mass for the
scalar field which acts as an infrared cutoff. Using such a
resummation scheme, we recently performed \cite{Wang} a 2-loop
calculation of $\rho_{\phi\phi}$. We found that for weak coupling
among the scalar fields ($\lambda/24 \ll 1$) the spectral function is
sharply peaked around the plasmon frequency
  \begin{equation}
  \label{freq}
  \omega_p({\bf k}) = \sqrt{{\bf k}^2+m_{\rm th}^2
                            - {{3m_{\rm th}^3}\over {\pi T}}}
                    \equiv \sqrt{{\bf k}^2+m_{\rm p}^2}
  \end{equation}
where
  \begin{equation}
  \label{mass}
   m_{\rm th}=T\sqrt{{\lambda\over 24}}
  \end{equation}
is the ``thermal mass" and $m_{\rm p}$ is the plasmon mass
\cite{Parwani}. The spectral function $\rho_{\phi\phi}$ can then be
expressed to good approximation in the form of a relativistic
Breit-Wigner function,
  \begin{equation}
  \label{rhobw}
    \rho_{\phi\phi}(k_0,{\bf k}) \approx {1\over \pi}\,
    { 2k_0\gamma({\bf k})\over
     (k_0^2 - \omega_p^2({\bf k}))^2
     + 4k_0^2\gamma^2({\bf k}) }\, ,
  \end{equation}
where
  \begin{equation}
  \label{gamma1}
    \gamma({\bf k}) =
    {{{\rm Im\,}\Sigma(\omega_p({\bf k}),{\bf k})}\over
      {2\omega_p({\bf k})}}
  \end{equation}
is the on-shell damping rate for the scalar plasmon. Its leading
contribution comes from the 2-loop diagram in Fig.~2 and is
given by \cite{Wang,Jeon2}
 \begin{eqnarray}
 \label{gammak}
    \gamma({\bf k})
    = {\lambda^2 T^2 \over
         {256\, \pi^3\,\omega_p({\bf k})}} \,
         {1\over |{\bf k}|}\int_0^{|{\bf k}|}
   &dq&
    \left[ L_2(\xi)+L_2\left({{\xi-\zeta}\over {\xi (1-\zeta)}}\right)
    \right.
  \nonumber\\
    &&\left. -L_2\left({\xi-\zeta \over 1-\zeta }\right)-
        L_2\left({{(\xi-\zeta)(1-\xi\zeta)}\over {\xi(1-\zeta)^2}}\right)
     \right] \, ,
 \end{eqnarray}
where
 \begin{equation}
 \label{xi}
    \xi=e^{-\beta\sqrt{{\bf k}^2+m_{\rm p}^2}}\, ,
    \qquad  \zeta=e^{-\beta\sqrt{{\bf q}^2+m_{\rm p}^2}} \, ,
 \end{equation}
and $L_2(z)$ is the Spence function
 \begin{equation}
 \label{spence}
   L_2(z) \equiv -\int_0^z dt\, {{\ln (1-t)}\over t}\, .
 \end{equation}
Substituting Eq.~(\ref{rhobw}) into Eq.~(\ref{eta2}) and integrating
over $k_0$ we obtain \cite{Hosoya,Jeon,Gangnus}
  \begin{equation}
  \label{eta3}
  \eta^{{\rm 1-loop}}_{\phi^2\phi^2} \approx {\beta\over 2\pi}
    \int{{d^3k}\over {(2\pi)^3}}\,
    {{f(\omega_p({\bf k})) \, [1+f(\omega_p({\bf k}))]}\over
      {\omega_p^2({\bf k}) \, \gamma({\bf k})}}\, .
  \end{equation}

\section{Results and discussion}\label{sec4}

We will now evaluate the momentum integral in (\ref{eta3}). It can be
studied analytically in the weak coupling limit $\lambda \ll 1$. Let
us first consider an approximation which has been investigated before
\cite{Ilyin,Jeon} where the momentum dependence of the plasmon width
is neglected and $\gamma({\bf k})$ is replaced by its zero-momentum
limit
 \begin{equation}
 \label{gamma0}
    \gamma(0) = { {\rm Im\,} \Sigma(m_{\rm p},0) \over 2\, m_{\rm p} }
    = {{\lambda^2\, T^2}\over {256\,\pi^3\, m_{\rm p}}}\,
      L_2 (e^{-m_{\rm p}/T})
    = { \lambda^2 T^2 \over 1536 \pi m_{\rm p}}\,
        \left[ 1 + {\cal O}(\sqrt{\lambda}\ln\lambda) \right] \, .
 \end{equation}
Inserting this into Eq.~(\ref{eta3}) and integrating by parts leads to
  \begin{equation}
  \label{eta3a}
    \left. \eta^{{\rm 1-loop}}_{\phi^2\phi^2}
    \right \vert_{\gamma({\bf k}) = \gamma(0)}
    = {384\, a^3 \over \pi^2\, T \, \lambda^2}
      \int_0^\infty {dx\over (x^2+a^2)^{3/2}} \,
      {1 \over e^{\sqrt{x^2+a^2}} - 1}\, ,
  \end{equation}
with $a=m_{\rm p}/T$. For small $a\ll 1$ the integral can be evaluated
following the methods of Appendix C in Ref.~\cite{Jackiw}, and we
obtain \cite{fn1}
  \begin{equation}
  \label{eta4}
    \left. \eta^{{\rm 1-loop}}_{\phi^2\phi^2}
    \right \vert_{\gamma({\bf k})=\gamma(0)}
    \approx {96\over {\pi T}}{1\over {\lambda^2}}
   \left( 1 - \sqrt{\lambda \over 6\pi^2} + {\cal O}(\lambda) \right)\, .
  \end{equation}

With the explicit momentum dependence \cite{Wang} of the plasmon decay
width from Eq.~(\ref{gammak}) this approximation can be avoided.
Inserting Eq.~(\ref{gammak}) into Eq.~(\ref{eta3}) and integrating by
parts, one finds
 \begin{equation}
 \label{etak}
    \eta^{{\rm 1-loop}}_{\phi^2\phi^2} = {64 \over \lambda^2 \, T}
    \int_0^\infty {dx \over e^{\sqrt{x^2+a^2}} - 1}\,
    {A(x;a) - x A'(x;a) \over A^2(x;a)}
 \end{equation}
where $x=k/T$, the prime denotes $d/dx$, and
 \begin{eqnarray}
 \label{A}
    A(x;a) & \equiv & {256\,\pi^3\over \lambda^2\, T^2}\,
      \omega_p({\bf k}) \, \gamma({\bf k})
    = {128\, \pi^3 \over \lambda^2\, T^2} \, {\rm Im\,} \Sigma
      (\omega_p({\bf k}), {\bf k})
 \\
    &=& {1\over x} \int_0^x dz\, \left[
         L_2(\xi) + L_2\left({{\xi-\zeta}\over {\xi (1-\zeta)}}\right)
        -L_2\left({{\xi-\zeta}\over {1-\zeta}}\right)
        -L_2\left({{(\xi-\zeta)(1-\xi\zeta)}\over {\xi(1-\zeta)^2}}\right)
     \right]
 \nonumber
 \end{eqnarray}
with (see Eq.~(\ref{xi}))
 \begin{equation}
 \label{xi1}
   \xi = e^{-\sqrt{x^2+a^2}}\, , \qquad
   \zeta = e^{-\sqrt{z^2+a^2}}\, , \quad {\rm and} \quad
   a = {m_{\rm p} \over T}\, .
 \end{equation}
The function $A(x;a)$ is shown in Fig.~3.  For large momenta,
$x\to \infty$, it approaches the constant value \cite{Jeon2}
 \begin{eqnarray}
 \label{a1}
   \lim_{x \to \infty} A(x;a)
   &=& -\int_0^\infty dz\, \ln\left( 1- e^{-\sqrt{z^2+a^2}} \right)
    = \int_a^\infty {\sqrt{\varepsilon^2 - a^2}\, d\varepsilon
                     \over e^\varepsilon - 1}
 \nonumber\\
   & \longrightarrow&
   \left\{
   \begin{array}{lcl}
   {\pi^2/6} & {\rm for} & a\to 0 \, ;\\
   a K_1(a)  & {\rm for} & a \gg 1 \, .
   \end{array}
   \right.
 \end{eqnarray}
At zero momentum it takes the value \cite{Jeon2}
 \begin{equation}
 \label{a2}
   A(0;a) = L_2(e^{-a})
 \end{equation}
which again approaches $\pi^2/6$ in the weak coupling limit $a =
\sqrt{\lambda/24} \to 0$. Unfortunately, in the weak coupling limit
the function $A(x;a)$ exhibits a very strong momentum dependence in
the region $a<x<1$; for $a=0$ the limit $x\to 0$ is non-analytic. If
we nonetheless neglect the momentum dependence of $A$,
Eq.~(\ref{etak}) becomes very simple:
 \begin{equation}
 \label{etak1}
    \eta^{{\rm 1-loop}}_{\phi^2\phi^2} \approx {64 \over \lambda^2 \, T}
    \, {6\over \pi^2}
    \int_0^\infty {dx \over e^{\sqrt{x^2+a^2}} - 1}\, .
 \end{equation}
In the limit $a\to 0$ this integral diverges logarithmically in the infrared.
The singular behaviour can be isolated by writing
 \begin{eqnarray}
 \label{logarithm}
   \int_0^\infty {dx \over e^{\sqrt{x^2+a^2}} - 1} &=&
   \int_a^\infty {\varepsilon  \over \sqrt{\varepsilon^2-a^2}}\,
          {d\varepsilon \over e^\varepsilon - 1} =
   \int_a^\infty {d\varepsilon \over e^\varepsilon - 1} +
   \int_a^\infty {\varepsilon -\sqrt{\varepsilon^2-a^2}
                  \over \sqrt{\varepsilon^2-a^2}}\,
          {d\varepsilon \over e^\varepsilon - 1}
 \nonumber\\
   &=& - \ln(1-e^{-a}) +
   \int_a^\delta {\varepsilon -\sqrt{\varepsilon^2-a^2}
                  \over \sqrt{\varepsilon^2-a^2}}\,
          {d\varepsilon \over \varepsilon}
   + {\cal O} \left({a^2\over \delta^2}\right)
 \nonumber\\
   &=& \ln\left({1\over a}\right) + \ln 2 + {\cal O}(a)
   + {\cal O} \left({a^2\over \delta^2}\right)\, .
 \end{eqnarray}
In the second line we cut off the second integral at the upper end at
a point $\delta$ with $a\ll \delta \ll 1$. This allows to approximate
the Bose distribution as $1/\varepsilon$. One easily verifies that the
remaining integral from $\delta$ to $\infty$ is finite and of order
$a^2/\delta^2$ as indicated. We thus find with this approximation
\cite{fn1}
 \begin{eqnarray}
 \label{etas}
   \eta^{{\rm 1-loop}}_{\phi^2\phi^2}
   &\approx& {384 \over \pi^2\, T}\,
   {1\over \lambda^2} \left( \ln \left({T\over m_{\rm p}} \right)
        +\ln 2 + {\cal O}\left({m_{\rm p}\over T}\right) \right)
 \nonumber\\
   &=& {192 \over \pi^2 \, T} \, {1 \over \lambda^2}
     \left( \ln \left( {1\over \lambda} \right) + \ln(96)
            + {\cal O}\left(\sqrt{\lambda}\right) \right)\, .
 \end{eqnarray}

The additional logarithmic divergence in the weak coupling limit of
(\ref{etas}) compared to (\ref{eta4}) is generic; its coefficient and
the next-to-leading constant term depend, however, on the momentum
dependence of $A(x;a)$. Due to the non-analytic behaviour of $A(x;0)$
near $x=0$ we were not successful in extracting an analytical
expression similar to (\ref{etas}) for the full integral (\ref{etak}).
The numerical result shown by the dots in Figure 4 can,
however, be excellently fit by
 \begin{equation}
 \label{etaratio}
   \eta^{{\rm 1-loop}}_{\phi^2\phi^2} \Big/
    \left. \eta^{{\rm 1-loop}}_{\phi^2\phi^2}
    \right \vert_{\gamma({\bf k})=\gamma(0)} =
    0.3282 \,\ln\left({T\over m_{\rm p}}\right) + 1.41682
    \approx {1\over \pi} \,\ln\left({T\over m_{\rm p}}\right) +
    \sqrt{2}
 \end{equation}
which suggests the analytical behaviour
 \begin{eqnarray}
 \label{etaratio1}
   \eta^{{\rm 1-loop}}_{\phi^2\phi^2} &=& {96 \over \pi^2 T}\,
   {1 \over \lambda^2} \left( \ln \left( {T\over m_{\rm p}} \right)
       + \pi\sqrt{2} + {\cal O}\left( {m_{\rm p}\over T} \right) \right)
 \nonumber\\
   &=& {48 \over \pi^2 \, T} \, {1 \over \lambda^2}
     \left( \ln \left( {1\over \lambda} \right) + \ln(24)
            + 2\pi\sqrt{2}
            + {\cal O}\left(\sqrt{\lambda}\right) \right)\, .
 \end{eqnarray}

As intuitively expected, the viscosity decreases as the coupling
strength increases: In relaxation time approximation all transport
coefficients are proportional to the relaxation time, which again is
inversely proportional to the scattering rate which grows with the
coupling strength. On the other hand, the rate at which our
$\eta_{\phi^2\phi^2}$ decreases with the coupling strength $\lambda$
is different from the behaviour of the physical shear viscosity
$\eta$. As discussed in the Introduction, the latter involves four
additional spatial derivatives acting on the scalar field. This
translates \cite{Hosoya} into an additional factor $k^4$ in the
integrand of Eq.~(\ref{eta3}), which removes the infrared divergence
of this integral in the limit $a\to 0$ and thereby also the leading
logarithmic term in our final result (\ref{etaratio1}). This
observation provides a partial explanation for the qualitatively
different behaviour of the viscosity as a function of $m_{\rm p}/T$
which was recently found by Jeon in a more complete study of $g\phi^3
+ \lambda \phi^4$ theory \cite{Jeon2}: his numerical results indicate
an increase of $\lambda^2 \eta$ with increasing $\lambda$.

Still, as first pointed out by Jeon in \cite{Jeon} and then
quantitatively analyzed in \cite{Jeon2}, the fact that the shear
viscosity $\eta$ is proportional to $1/\lambda^2$ raises a serious
problem: simple power counting arguments \cite{Jeon} show that then
all planar ladder diagrams of the type shown in Fig.~5 can also
contribute to the leading order result for the viscosity. The
summation of this infinite series of ladder diagrams is nontrivial and
has been recently performed by Jeon~\cite{Jeon2} using the imaginary
time formalism.  This work was a genuine {\em tour de force}, and the
prospect of generalizing it to non-abelian gauge theories like QCD
seems frightening. However, based on the results presented in this
work, we believe that the methods developed here in the framework of
TFD will help to redo the analysis in a more efficient way and allow
for an easier generalization to QCD. Work in this direction is in
progress.

\acknowledgments

Two of authors (E.W. and X.Z.) are grateful to Liu Lianshou, Li
Jiarong and R.D. Pisarski for helpful discussions. This work was
supported in part by the Deutsche Forschungsgemeinschaft (DFG), the
Bundesministerium f\"ur Bildung und Forschung (BMBF), the National
Natural Science Foundation of China (NSFC) and the Gesellschaft f\"ur
Schwerionenforschung (GSI).


 \begin{figure}
\epsfxsize=10cm
\centerline{\epsfbox{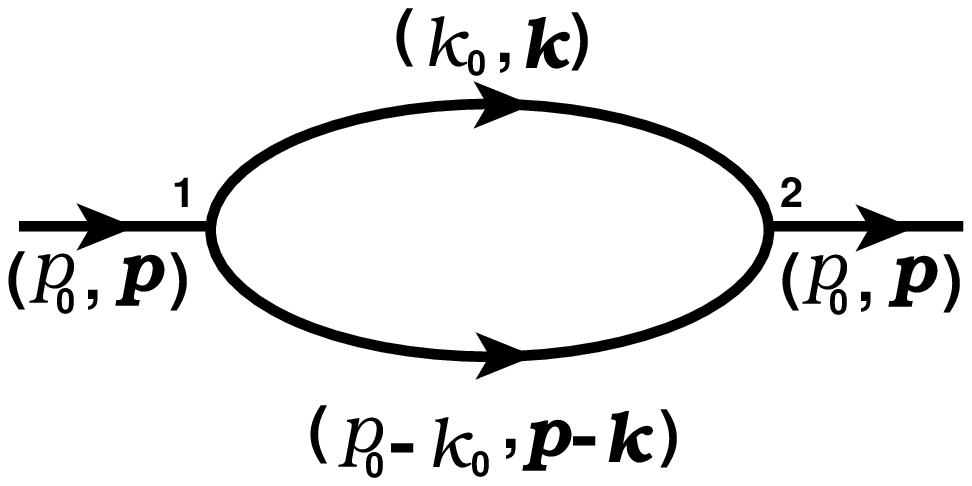}}
\vskip 0.4cm
 \caption{One-loop skeleton diagram for $\Delta^{12}_{ \phi^2 \phi^2 }
    (p_0, {\bf p})$. The heavy lines denote full single particle
    propagators.}
 \label{F1}
 \end{figure}

 \begin{figure}
\epsfxsize=10cm
\centerline{\epsfbox{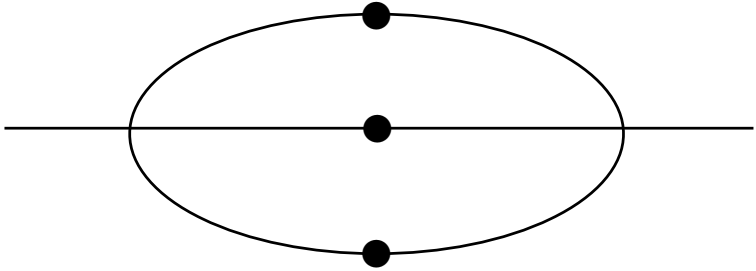}}
\vskip 0.4cm
 \caption{The leading contribution to the imaginary part of self energy
     ${\rm Im\,}\Sigma$. The dotted lines denote resummed
     effective propagators.}
 \label{F2}
 \end{figure}\newpage

 \begin{figure}
\epsfxsize=10cm
\centerline{\epsfbox{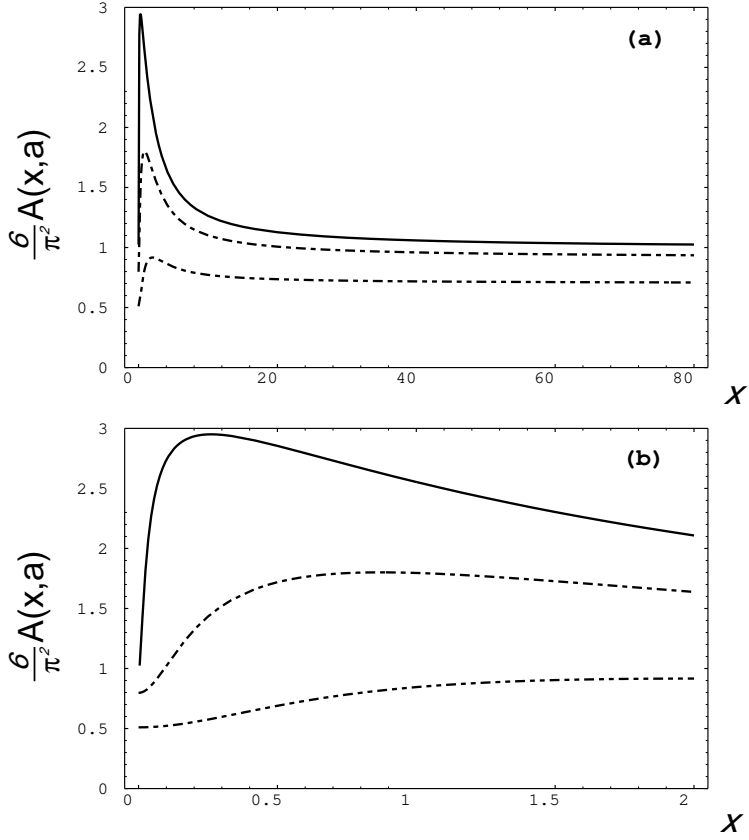}}
\vskip 0.4cm
 \caption{(a) The function $(6/\pi^2) A(x;a)$ vs. $x$ for $a$ = 0.01,
    0.1, and 0.4 (from top to bottom). (b) Close-up of (a) for small
    $x$.}
 \label{F3}
 \end{figure}

 \begin{figure}
\epsfxsize=10cm
\centerline{\epsfbox{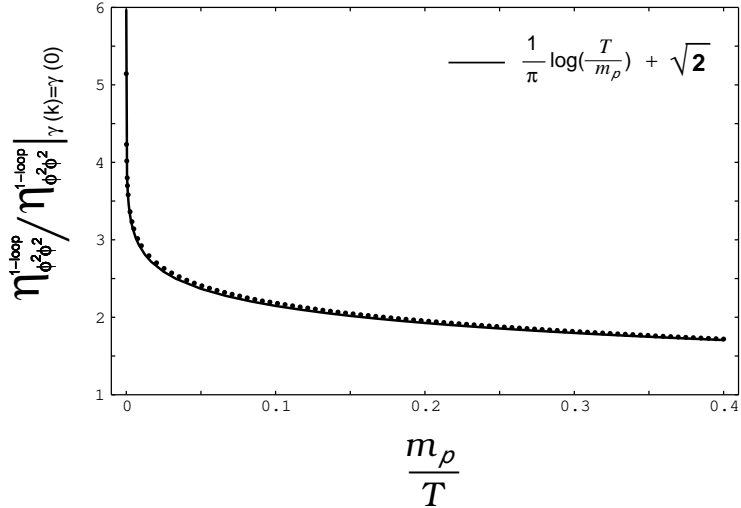}}
\vskip 0.4cm
 \caption{The dots represent numerical results for the ratio between
   $\eta^{\rm 1-loop}_{\phi^2\phi^2}$ and $\eta^{\rm
   1-loop} _{\phi^2 \phi^2} \Big\vert_{\gamma({\bf k})=\gamma(0)}$ as
   a function of $m_{\rm p}/T$. The solid line indicates a fit of the
   form $(1/\pi)\ln(T/m_{\rm p})+\protect\sqrt{2}$.}
 \label{F4}
 \end{figure}

 \begin{figure}
\epsfxsize=10cm
\centerline{\epsfbox{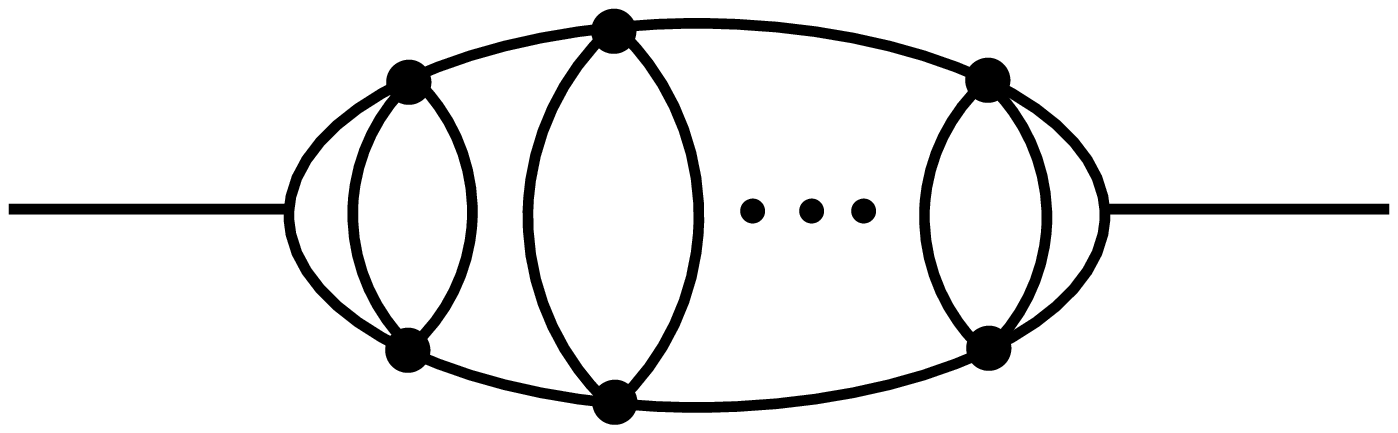}}
\vskip 0.4cm
 \caption{The multi-loop planar-ladder skeleton diagram in $\lambda\phi^4$
     theory.}
 \label{F5}
 \end{figure}


\begin{thebibliography}{99}
\bibitem{fn1}
   Our result should be compared with Eq.~(5.12) in Ref.~\cite{Jeon}.
   If in that expression the mass is replaced by the thermal mass
   $m_{\rm th}$ or $m_{\rm p}$, it exhibits a logarithmic dependence
   $\sim \lambda^{-2} \ln (a \lambda)$ on the coupling constant.  In
   \cite{Jeon} it is stated that this result (with $a=2$) is obtained
   if the momentum dependence of the plasmon {\em width} is neglected;
   this disagrees with our result (\ref{eta4}) in that limit. Jeon's
   result (5.12) with $a=2$ is, however, recovered if the momentum
   dependence of the imaginary part of the plasmon's {\em self energy}
   is neglected, see our Eq.~(\ref{etas}).
\end{thebibliography}
\end{document}